%% file: main.tex
\documentclass[letterpaper]{article} %
\usepackage{aaai24}  %
\usepackage{times}  %
\usepackage{helvet}  %
\usepackage{courier}  %
\usepackage[hyphens]{url}  %
\usepackage{graphicx} %
\usepackage{natbib}  %
\usepackage{caption} %
\usepackage{algorithm}
\usepackage{algorithmic}

\usepackage{newfloat}
\usepackage{listings}
\DeclareCaptionStyle{ruled}{labelfont=normalfont,labelsep=colon,strut=off} %
\floatstyle{ruled}
\newfloat{listing}{tb}{lst}{}
\floatname{listing}{Listing}
\usepackage{subcaption}

\usepackage{todonotes}

\usepackage{tikz}
\usepackage{pgfplots}
\pgfplotsset{compat=1.18}
\usetikzlibrary{calc}

\usepackage{xspace}
\newcommand\KeY{Ke\kern-0.75ptY\xspace}

\newcommand{\trueSym}{\ensuremath{\texttt{true}}}
\newcommand{\falseSym}{\ensuremath{\texttt{false}}}

\lstdefinestyle{mystyle}{
    keywordstyle=\bfseries,
    numberstyle=\small,
    basicstyle=\ttfamily\scriptsize,%
    escapeinside={\%*}{\%},
    breakatwhitespace=false,
    breaklines=true,
    captionpos=b,
    keepspaces=true,
    numbers=left,
    numbersep=5pt,
    showspaces=false,
    showstringspaces=false,
    showtabs=false,
    tabsize=2,
    frame=single,
    morekeywords={assert,assume,false,true,bool},
}

\usepackage{amsthm}
\usepackage{amsmath}
\usepackage{amssymb}
\usepackage{mathtools}

\newtheorem{definition}{Definition}
\newtheorem{lemma}{Lemma}

\newtheorem{example}{Example}

\newif\ifAAAI
\AAAIfalse

\usepackage[createShortEnv,conf={restate,text proof translated={Proof of }}]{proof-at-the-end}

\usepackage{paralist}
\usepackage{cleveref}

\title{An Information-Flow Perspective on Algorithmic Fairness\thanks{Please consider the Corrigendum on \cpageref{corrigendum}}}
\author{
    Samuel Teuber\textsuperscript{\rm 1},
    Bernhard Beckert\textsuperscript{\rm 1}
}
\affiliations{
    \textsuperscript{\rm 1}Karlsruhe Institute of Technology\\

    Am Fasanengarten 5\\
    76131 Karlsruhe\\
    \{teuber,beckert\}@kit.edu
}

\begin{document}

\maketitle

\begin{abstract}
This work presents insights gained by investigating the relationship between
algorithmic fairness and the concept of secure information flow. The problem
of enforcing secure information flow is well-studied in the context of
information security: If secret information may ``flow'' through an algorithm
or program in such a way that it can influence the program’s output, then
that is considered insecure information flow as attackers could potentially
observe (parts of) the secret.
There is a strong correspondence between secure information flow
and algorithmic fairness: if protected attributes such as race, gender, or age
are treated as secret program inputs, then secure information flow means that
these ``secret'' attributes cannot influence the result of a program.
While most research in algorithmic fairness evaluation concentrates on
studying the impact of algorithms (often treating the algorithm as a black-box),
the concepts derived from information flow can be used both
for the analysis of \emph{disparate treatment} as well as \emph{disparate impact} w.r.t.\ a structural causal model.

In this paper, we examine the relationship between quantitative as well as
qualitative information-flow properties and fairness. Moreover, based on this duality,
we derive a new quantitative notion of fairness called \emph{fairness spread},
which can be easily analyzed using quantitative information flow and which strongly relates to counterfactual fairness.
We demonstrate that off-the-shelf tools for information-flow properties can be
used in order to formally analyze a program's algorithmic fairness properties,
including the new notion of fairness spread as well as established notions
such as demographic parity.
\end{abstract}

\section{Introduction}
The problem of enforcing secure information flow is well-studied in the context of information security
and allows to rigorously analyze whether secret information can ``flow'' through an algorithm potentially influencing its result.
A wide spectrum of methods and tools is available to analyze and quantify to what extent a given program satisfies
information-flow properties~\cite{Lampson73,denning1975secure,Denning76,DarvasHS05,Smith2009,beckert_information_2013,klebanov_precise_2014,KeYBook,biondi_scalable_2018}.
This paper investigates the relationship between algorithmic fairness and information flow by building upon our initial findings from \cite{beckert_algorithmic_2022,TeuberEWAF23}.
\ifAAAI
An extended version of this paper can be found on arXiv~\cite{AAAI24arxiv}.
\fi
We demonstrate that there is a strong correspondence between qualitative as well as quantitative information-flow properties
and algorithmic fairness: if protected attributes such as race, gender, or age
are treated as secret program inputs, then secure information flow means that
these ``secret'' attributes cannot influence the result of a program.
Our approach is thus a white-box analysis of the algorithm and analyzes the decision-making \textit{process} rather than only it's outcome.
This allows us to find disparities in treatment even for cases where these disparities are not immediately observable in a statistical evaluation of the algorithm's outcome.

A large body of research, see e.g.\ \cite{Beutel2019,taskesen_statistical_2021,hardt_equality_2016,mitchell2018predictionbased}, studies the disparate impact of algorithms using statistical metrics such as Equalized Odds or Equality of Opportunity~\cite{hardt_equality_2016}, Demographic Parity or certain forms of Individual Fairness.
Many of these approaches consider the algorithm as a black-box, as they do not take into account the algorithm's decision mechanism, but only its outcome.
While there exist some approaches for the white-box analysis of algorithms, e.g.\ \cite{Albarghouthi2017,Ghosh2021a,Ghosh_Basu_Meel_2022,bastani2019probabilistic}, they are  fairness-specific and do not consider the relationship between information flow and fairness discussed here.
Another direction of related research analyzed information flow on the level of software architecture~\cite{ramadan_model-based_2018,ramadan_explaining_2019} rather than program-level to mitigate fairness risks.
That research is in some ways orthogonal to our work as it does not consider program code analysis and furthermore does not investigate the concrete correspondence between information flow and algorithmic fairness presented here.

The structure of this paper is as follows:
After providing the necessary background on program analysis and causality (\Cref{sec:background}), we demonstrate and discuss the relationship between multiple flavors of qualitative information flow and demographic parity -- including options to allow for certain justified exceptions to fairness guarantees (\Cref{sec:qualitative-information-flow}).
\Cref{sec:quantative-information-flow} then extends the relation between information flow and fairness from the qualitative into the quantitative world.
To this end, we derive a new quantitative notion of fairness called \emph{Fairness Spread}, which can be computed using quantitative information-flow analysis tools.
In \Cref{sec:qualitative-information-flow,sec:quantative-information-flow} we assume the statistical independence of protected group attributes and unprotected attributes which initially limits our approach to disparate treatment alone.
\Cref{sec:proxy} demonstrates how this limitation can be lifted in the qualitative and quantitative approach by integrating causal models into our approach.
Finally, \Cref{sec:examples} demonstrates the applicability of our approach on two case studies.

Throughout this paper, we will demonstrate our ideas using the three exemplary programs shown in \Cref{lst:example_algos}:
All programs make a loan allocation decision based on a protected group attribute (\texttt{group}) and an unprotected variable containing the credit score (\texttt{score}).
As group attribute, we assume a protected category such as gender, rage or age.
The algorithms vary in their use of the two variables: While \Cref{lst:group_only} is clearly biased as its decision is solely dependent on the group attribute, the other two decision algorithms use the credit score as well (\Cref{lst:mixed}) or exclusively (\Cref{lst:score_only}).
There exist a variety of off-the-shelf tools for the analysis of programs w.r.t.\ information-flow properties.
For example, we will be using the interactive theorem prover \KeY~\cite{ahrendt2005key,KeYBook}, which features a (partially) automated analysis technique for qualitative information-flow properties, the automated information-flow analysis tool Joana~\cite{tooljoana2013atps,joana14it} and the tool counterSharp~\cite{teuber_quantifying_2021} which allows the sound, bit-precise quantiative analysis of C programs using propositional model counting.

\begin{figure}[t]
\centering
\begin{minipage}[t]{0.48\linewidth}
\begin{subfigure}[t]{\linewidth}
\begin{lstlisting}[language=Python,style=mystyle,escapechar=|]
def c1(group, score):
  return (group != 0)
\end{lstlisting}
\caption{Group Decision}
\label{lst:group_only}
\end{subfigure}

\vspace{1em}
\begin{subfigure}[t]{\linewidth}
\begin{lstlisting}[language=Python,style=mystyle,escapechar=\%]
def c2(group, score):
  return (score>=8)
\end{lstlisting}
\caption{Score Decision}
\label{lst:score_only}
\end{subfigure}
\end{minipage}
\begin{subfigure}[t]{.48\linewidth}
\begin{lstlisting}[language=Python,style=mystyle,escapechar=\%]
def c3(group, score):
  if (group >= 6):
    return (score >= 8)
  else:
    return (score >= 6)
\end{lstlisting}
\caption{Mixed Decision}
\label{lst:mixed}
\end{subfigure}

\caption{Three exemplary credit decision algorithms}
\label{lst:example_algos}
\end{figure}

\section{Preliminaries}
\label{sec:background}
\newcommand{\statespace}{\ensuremath{\mathcal{S}}}
\newcommand{\project}[2]{\ensuremath{#1\!\downharpoonright_{#2}}}

\paragraph{Program Analysis}
This work is concerned with the analysis of decision procedures or more specifically with the analysis of program code.
As the algorithm's inputs, we consider a protected group attribute $g \in \mathcal{G}$ and an independent, unprotected input $u \in \mathcal{U}$.
The output of the algorithm is a decision $d \in \mathcal{D}$ (e.g., whether a person is awarded a credit or not).
For a given group and unprotected attribute, we will usually describe the algorithm as a deterministic function $P$ that returns a decision $d=P\left(g,u\right)$.
Note, that this restriction is not necessarily relevant for practical use:
A probabilistic algorithm can be turned into a deterministic algorithm by adjusting the domain of the unprotected attribute~$\mathcal{U}$.
We assume $\mathcal{G}$ and $\mathcal{U}$ to be finite which is a reasonable assumption for realistic programs: Even for the case where inputs are technically continuous they can be quantized with a user-defined precision.

We now consider the random variables $G \in \mathcal{G}$ and $U \in \mathcal{U}$, which we assume to be independent.
\Cref{sec:proxy} will demonstrate how we can extend our approach to model correlations between unprotected and protected variables.
By entering $G$ and $U$ into $P$, we obtain the random variable $D = P\left(G,U\right)$ which is dependent on the function~$P$ as well as $G$ and~$U$.
The probability $\Pr\left[D=d\right]$ that $D$ takes on a certain value $d \in \mathcal{D}$ is then given by:
\[
{{\textstyle\sum_{\left(g,u\right) \in \mathcal{G}\times\mathcal{U}}}
\Pr\left[P\left(g,u\right)=d\right]
\Pr\left[G=g\right] \Pr\left[U=u\right]}.
\]

By analyzing the (conditional) distribution of the random variable $D$, we are able to analyze properties such as the probability that an individual of group $g \in \mathcal{G}$ obtains an outcome $d \in \mathcal{D}$.

\paragraph{Causal Models}
\Cref{sec:proxy} will use causal models and counterfactuals to extend our approach to dependent distributions of $G$ and $U$.
To this end, we briefly introduce causal models and the notion of counterfactual fairness following the presentations of \cite{pearl2009causal} and \cite{kusner2017counterfactual}:
\begin{definition}[Causal Model]
    We define a \emph{causal model} $C=\left(V,U,F\right)$ to consist of a set $V=\left\{V_1,\dots,V_n\right\}$ of modeled variables, a set $B$ of
    latent background variables for which we assume no causal mechanism, and a set $\left\{f_1,\dots,f_n\right\}$ of functions defining the variables $V_i$, i.e. $V_i=f_i\left(\mathrm{pa}_i,B_{\textrm{pa}_i}\right)$ where $\mathrm{pa}_i \subseteq V \setminus \left\{V_i\right\}$ and $B_{\textrm{pa}_i} \subseteq B$.
\end{definition}
Here, the variables in $\mathrm{pa}_i$ are the \emph{parents} of $V_i$ and we assume that the computation of all the $V_i$ from background variables in $B$ has the structure of a directed acyclic graph.
By abuse of notation, we will sometimes refer to $B$ as a single random variable.
\newcommand{\intervene}[3]{\ensuremath{#1_{#2 \leftarrow #3}}}
Given a causal model $C$, a variable $V_i$ and a value $x$, we can perform an \emph{intervention} $\intervene{C}{V_i}{x}$ resulting in a modified causal model $C'$ that is the same as $C$ except for $f_i$ which is set to $f_i=x$.
For our purposes, we can interpret $C$ as a deterministic program computing all $V_i \in V$ when provided with values for the background variables $B$.
Given a decision procedure $P\left(G,U\right)$ and a causal model $C$ containing variables $G,U$ we denote as $\hat{P}_C\left(B\right)$ the program which takes $B$ as input, computes $G$ and $U$ through $C$ and then returns $P\left(G,U\right)$.
Moreover, we denote with $\hat{P}_C\left(g,B\right)$  the program which computes $G$ and $U$ through $\intervene{C}{G}{g}$ and returns the result of $P$.
Using this notation, we can now define counterfactual fairness as follows:
\begin{definition}[Counterfactual Fairness]
A program $P$ with inputs $G$ and $U$ is counterfactually fair with respect to a causal model $C$ iff, for any \mbox{$g_1,g_2 \in \mathcal{G}, u \in \mathcal{U}, d \in \mathcal{D}$}, it holds that:
$\Pr\left[\hat{P}_C(g_1,B) = d \;\middle|\; U=u,G=g_1 \right]
=\Pr\left[\hat{P}_C(g_2,B) = d \;\middle|\; U=u,G=g_1 \right]
$
\end{definition}
It is worth noting, that in this scenario we condition on $G=g_1$ in \emph{both} cases as $B$ must be conditioned w.r.t. to the factual outcome (not the counterfactual).

\section{Qualitative Information Flow and Demographic Parity}
\label{sec:qualitative-information-flow}
Information Flow is a well-studied concept originally introduced for software security analysis, see, e.g., \cite{Lampson73,denning1975secure,DarvasHS05,Smith2009,beckert_information_2013,klebanov_precise_2014,KeYBook,biondi_scalable_2018}.
Consider a program $P$ for checking whether a user entered the right password.
Such a program will have two inputs: The correct password and the password entered by the user.
Ideally, we would want that such a program will not leak any information about the correct password.
An information-flow analysis checks whether information about the secret password is leaked by the program resp.\ how much information is leaked.
A ``bad'' program that, for example, returns the full secret password to the user would be said to contain insecure information flow.
To this end, we distinguish between variables with \emph{high} security status (e.g., the correct password) and variables with \emph{low} security status (e.g., the user's input and the program's return value).
In its most basic form, insecure information flow is defined as follows: a program contains insecure information flow iff there exist two high security status values $h,h'$ and a low security status value $l$ such that the program returns different outputs for the inputs $\left(h,l\right)$ and $\left(h',l\right)$.
That definition is sufficient for many applications. But, even a ``good'' password checker will leak some information about the password and have insecure information flow in the above sense because the output depends on whether the entered password is correct and, thus, depends on the secret.
Therefore, there are various refined flavors of information flow that take into account such considerations.
From a formal methods perspective, information flow is a \emph{hyperproperty}:
Where regular program properties talk about a single program run and analyze whether it satisfies a given specification, hyperproperties make statements about multiple program runs and their releation to each other.
Due to the large research body on information flow, there exist many tools for analysing information-flow hyperproperties.

In the following, we demonstrate that information flow provides an interesting framework for the analysis of Algorithmic Fairness.
We analyze what information about a protected group variable is leaked to the low security decision variable.
Thus, the group variable $G$ is assigned high security status while the unprotected variable $U$ as well as the decision $D$ are assigned low security status.
A crucial assumption in this as well as the following section is the independence of the random variable describing the group $G \in \mathcal{G}$ and the random variable describing unprotected properties $U \in \mathcal{U}$.
In \Cref{sec:proxy}, we demonstrate that we are nonetheless able to analyze the case of correlated variables as well when provided with a model of their dependence.

\subsection{Unconditional Information Flow and Fairness}
We first define the basic information flow property.
We call this property \emph{Unconditional Noninterference} and it represents the strictest variant of secure information flow:
\begin{definition}[Unconditional Noninterference, see e.g.~\cite{KeYBook}]
A program $P$ satisfies \emph{unconditional noninterference} iff for all $g_1,g_2 \in \mathcal{G}$ and for all $u \in \mathcal{U}$ it holds that $P\left(g_1,u\right) = P\left(g_2,u\right)$.
\end{definition}
This definition forbids any and all information flow between $G$ and the outcome $D$ of the program.
The notion of Unconditional Noninterference is similar to fairness through unawareness~\cite{grgic2016case}. However, our approach decidedly allows the program~$P$ to \emph{process} an input from~$\mathcal{G}$ as long as it does not influence the outcome (see, e.g., the example in~\Cref{subsec:case_study:tax}).
There is also a correspondence of Unconditional Noninterference to counterfactual fairness~\cite{kusner2017counterfactual}, but that concept typically uses the rich framework of causality to reason about the relations between $G$ and $U$ -- a relation we will intentionally only discuss in \Cref{sec:proxy} in order to distinguish disparate treatment and impact.
There is furthermore a close relation to Demographic Parity~\cite{DworkHPRZ12,mitchell2018predictionbased}:
\begin{lemmaE}[Unconditional Noninterference implies Demographic Parity]%
    \label{lem:uncond_dem_parity}
    Let $P$ be a program and let $G,U$ be distributed arbitrarily, but independently.
    If $P$ satisfies unconditional noninterference, then the outcome of $P$ satisfies \emph{demographic parity}, i.e., for all $d \in \mathcal{D}, g_1,g_2 \in \mathcal{G}$ it holds that
    $
    \Pr\left[P\left(G,U\right)=d \middle| G=g_1\right]=\Pr\left[P\left(G,U\right)=d \middle| G=g_2\right]
    $
\end{lemmaE}
\begin{proofE}
For all $g_1,g_2 \in \mathcal{G}$ and $d \in \mathcal{D}$:
\begin{align*}
&\Pr\left[P\left(g_1,U\right) = d\right]\\
=\enspace&\sum_{u \in \mathcal{U}} \Pr\left[P\left(g_1,U\right)=d\mid U = u\right]\Pr\left[U=u\right]\\
=\enspace&\sum_{u \in \mathcal{U}} \Pr\left[P\left(g_1,u\right)=d \mid U=u\right]\Pr\left[U=u\right]\\
\intertext{By the definition of unconditional noninterference we can exchange $P(g_1,u)$ for $P(g_2,u)$:}
=\enspace&\sum_{u \in \mathcal{U}} \Pr\left[P\left(g_2,u\right)=d\mid U=u\right]\Pr\left[U=u\right]\\
=\enspace& \Pr\left[P\left(g_2,U\right) = d\right]
\end{align*}
\end{proofE}
We now analyze the example programs shown in \Cref{lst:example_algos} with respect to Unconditional Noninterference:
For the case of \Cref{lst:group_only}, we observe that for all $u$ it holds that $\texttt{c1}\left(0,u\right) \neq \texttt{c1}\left(1,u\right)$.
Thus, this first program as well as \Cref{lst:mixed} do not satisfy unconditional noninterference -- contrary to the program in \Cref{lst:score_only} which ensures demographic parity.

The great benefit of the correspondence between information flow and fairness is that we can use information-flow tools for fairness analyses. Accordingly, we can use methods such as the program verification tool~\KeY~\cite{KeYBook} to prove that the program in \Cref{lst:score_only} satisfies unconditional noninterference and, thus, demographic parity.

Note, that unconditional noninterference is a strictly stronger property than demographic parity and that there may be cases where an algorithm satisfying demographic parity does not satisfy noninterference%
\ifAAAI%
:
\else%
.
For example, for $\mathcal{D}=\left\{1,2\right\}$ and $\mathcal{G}=\left\{1,2\right\}$ both uniformly distributed, a program $P$ returning \texttt{true} iff $\left(g,u\right)$ are $\left(1,1\right)$ or $\left(2,2\right)$ will satisfy demographic parity, but not unconditional noninterference since the output depends on the value of $g$.
\fi
When disparate treatment does not lead to disparate impact in terms of the binary outcome, a program not satisfying unconditional noninterference may still yield demographic parity in its results and the disparity will initially be invisible when analyzed statistically.
At the same time, a program could equally be considered unfair in this case due to the decision making process that uses a protected attribute.
\Cref{apx:further_examples} demonstrates this on a concrete example.

In practice, it will not always be possible to conform to the strict standard of unconditional noninterference.
For example, in the case of a password checker we will have to leak the information whether or not the user entered the right password which would violate unconditional noninterference.
Similarly, there may be constraints which make it impossible to conform to unconditional noninterference in the fairness scenario.
This raises the question in which ways one can relax this property while still retaining certain fairness guarantees.
To this end, we can proceed in three manners:
\begin{enumerate}
    \item We provide \textit{restricted categories}, i.e., subgroups of all individuals, %
    according to which an algorithm may discriminate while ensuring that discrimination only happens along these categories (restricted information flow, see \Cref{sec:restricted-information-flow}).
    \item We provide conditions under which a \textit{declassification} of group attributes is allowed for cases where this is morally and/or legally justified (conditional information flow, see~\Cref{apx:conditional-information-flow}).
    \item We analyze information flow as a quantitative rather than a qualitative property (quantitative information flow, see \Cref{sec:quantative-information-flow}).
\end{enumerate}

\subsection{Restricted Information Flow and Fairness}
\label{sec:restricted-information-flow}
To make the allowed types of discrimination explicit, one can rely on the notion of \emph{Restricted Information Flow} where we provide categories within which all individuals must be treated equally.
To this end, we introduce a \emph{Restricted Classification}:%
\begin{definition}[Restricted Classification]%
    A \emph{Restricted Classification} is a function $R: \mathcal{G} \times \mathcal{U} \to \mathcal{R}$ mapping each combination $\left(g,u\right)$ to a class $r$ of a finite set $\mathcal{R}$.
\end{definition}%
Based on this notion of restricted classification, we can introduce \emph{Restricted Information Flow}:
\begin{definition}[Restricted Information Flow]
Let $R$  be some restricted classification.
A program $P$ satisfies \emph{Restricted Information Flow} for $R$ iff, for all $g_1,g_2 \in \mathcal{G}$ and all $u \in \mathcal{U}$:
\[R\left(g_1,u\right)=R\left(g_2,u\right) \enspace\mbox{implies}\enspace P\left(g_1,u\right) = P\left(g_2,u\right) \enspace.\]
\end{definition}
We can relate this notion of information flow with a conditional variant of demographic parity:
\begin{lemmaE}
    \label{lem:restricted_information_flow}
    Let $P$ be a program and $G,U$ be distributed arbitrarily, but independently.
    If $P$ satisfies Restricted Information Flow for a restricted class $R$, then, for each $r \in \mathcal{R}$, the outcome of $P$ satisfies conditional demographic parity with respect to the condition $R\left(G,U\right)=r$, i.e., for all $r \in \mathcal{R}$ and all $d \in \mathcal{D}, g_1,g_2 \in \mathcal{G}$:
    \begin{align*}
    \Pr\left[P\left(G,U\right)=d \mid G=g_1, R\left(G,U\right)=r \right]\\
    =\enspace\Pr\left[P\left(G,U\right)=d \mid G=g_2, R\left(G,U\right)=r \right]
    \end{align*}
\end{lemmaE}
\begin{proofE}
 Restricted Information Flow for a given $R$ implies that, for each $r \in \mathcal{R}$, we have Conditional Information Flow for the condition $R\left(G,U\right)=r$.
 \Cref{lem:conditional_dem_parity} then implies conditional demographic parity for each condition $R\left(G,U\right)=r$.
\end{proofE}
For example, for the program in \Cref{lst:mixed}, we can define a restricted classification $R\left(g,u\right) : \mathcal{G}\times\mathcal{U} \to \left\{\trueSym,\falseSym\right\}$ which maps to categories as follows: $ \left(g,u\right) \mapsto \left(6 \leq u \land u < 8 \land g \geq 6\right)$.
Then, Restricted Information Flow holds for \Cref{lst:mixed} with respect to~$R$.
A possible intuition behind this restriction is the idea that when an individual close to retirement age applies for a loan, it may be a justified disparate treatment that the loan is only granted if the individual can afford higher credit rates to ensure a timely redemption %
before retirement --- which is only possible for individuals with a higher credit score.
Thus, it may be justified to consider an individual's group (in this case their age in decades) if their credit score is in between the regular threshold and the threshold for near-retirement loans\footnote{This is meant as an example and does not make a claim as to whether or not this practice is indeed legally or morally justified.}.
We can analyze and prove the Restricted Information Flow property using the interactive theorem prover \KeY.
Restricted information flow makes the limitations of a fairness guarantee explicit and thus allows an audit of the classification along which individuals are disparately treated.
Providing auditors with the classification function $R$ can represent the starting point of a debate on whether the classification is justified and chosen correctly.

In cases where we only require a fairness guarantee limited to a subset of individuals, we place all individuals of that subset into one category while placing all other individuals into singleton categories each only containing a single possible input.
A more intuitive version of this approach (\emph{Conditional Information Flow}) is discussed in \Cref{apx:conditional-information-flow}.

\section{Quantitative Information Flow and Fairness}
\label{sec:quantative-information-flow}
Quantitative Information Flow, see e.g.~\cite{Smith2009}, transforms the qualitative security analysis of software into a metric for a software's security.
A good quantitative metric for information flow will satisfy two requirements:
\begin{inparaenum}
    \item Its minimum (optimal value) is reached if the program satisfies unconditional noninterference.
    \item A less secure program will have a higher value of the metric.
\end{inparaenum}
Quantitative Information Flow can be used to compare different program variants (to analyze which version is more secure) or to provide a bound on the likelihood of a breach.
One well-established technique for the quantitative analysis of information flow is the metric of \emph{min-entropy}~\cite{Smith2009}.
Its value is based on the concept of \emph{Vulnerability}, which corresponds to the probability that an attacker who observes the low inputs and the low output can correctly guess a variable of high security status in a single try.
Transferring that notion to the context of fairness, we define Conditional Vulnerability as follows:
\begin{definition}[Conditional Vulnerability]
    \label{def:vulnerability}
    For a program $P$ and random independent variables $G,U$, we define the \emph{Conditional Vulnerability}~$V\left(G \mid P\left(G,U\right),U\right)$ of $G$ w.r.t.\ $P\left(G,U\right)$ and $U$ as follows:
    \begin{align*}
      \sum_{u \in \mathcal{U}} \sum_{d \in \mathcal{D}} \max_{g \in \mathcal{G}}
      \big(&\Pr\left[P\left(G,U\right) = d,U=u\right] \cdot \\
        &\quad\Pr\left[G=g \mid P\left(G,U\right) = d,U=u\right]\big)\\
    \end{align*}
\end{definition}

We demonstrate this definition on a small example from the security context:
Consider a password checker for a correct password $G \in \mathcal{G}=\left\{1,2,3\right\}=\mathcal{U}$. The program returns $1$ if the entered password $U \in \mathcal{U}$ was correct and $0$ otherwise.
If an attacker enters a password $U \in \mathcal{U}$, there are two possibilities: Either the password is correct in which case $P$ returns $1$ and the attacker now knows the secret $G$ or the password is incorrect and the attacker has a 50\% chance of guessing the password correctly.
Thus, the Conditional Vulnerability of $G$ with respect to $P\left(G,U\right)$ and uniform $U$, i.e., the likelihood of an attacker guessing the correct password after 1 try, is:%
$1/3*1 + 2/3*1/2=2/3$. %
Conditional Vulnerability can be easily computed using off-the-shelf tools (see \Cref{apx:compute_cond_vuln}).
Below, we discuss whether Conditional Vulnerability is indeed a suitable metric for an algorithm's fairness.
We consider this question for binary decisions, i.e., $\left|\mathcal{D}\right|=2$.
Similarly as for security, there are two requirements that a fairness metric should satisfy:
\begin{inparaenum}
    \item The metric returns its minimum (optimal) value if the program satisfies (un)conditional demographic parity;
    \item A less fair program has a higher value of the metric.
\end{inparaenum}

\subsection{Naive Approach}
Unfortunately, a naive approach where Conditional Vulnerability is used as a fairness metric does not satisfy the above two requirements.
This is, because  $V\left(G \mid P\left(G,U\right), U\right)$ measures two properties at the same time:
On the one hand, we measure how much information a program leaks about the protected variable $G$.
On the other hand, we also measure how easy it is to guess $G$ based on its probability distribution.
Thus, for Conditional Vulnerability, it does not make a difference whether one can easily guess $G$ due to information leakage in the program or due to the distribution over~$G$.
To this end, consider the following example:
\begin{example}[Arbitrarily unfair program for constant Conditional Vulnerability]
\label{example:qif:unfair}
Let there be two groups $\mathcal{G}=\left\{0,1\right\}$.
For simplicity, we assume $\mathcal{U}=\left\{0\right\}$ for the unprotected attribute.
As before, we consider two possible outcomes, i.e.,~$\mathcal{D}=\left\{0,1\right\}$.
Let $1$ be an advantaged majority group and let $\mu_1 = \Pr\left[G=1\right]$.
We now consider the value of $V\left(G \mid P\left(G,U\right),U\right)$ which can be computed as:
$        \max_{g \in \mathcal{G}}  \left(
            \Pr\left[G=g\right]\cdot\Pr\left[P\left(G,0\right) = 0 \mid G=g\right]
        \right)%
        +\\
        \quad\max_{g \in \mathcal{G}}  \left(
            \Pr\left[G=g\right]\cdot\Pr\left[P\left(G,0\right) = 1 \mid G=g\right]
        \right)
$.
Consider the case where $\min_{d \in \mathcal{D}}\Pr\left[P\left(1,0\right)=d\right]\geq \frac{1-\mu_1}{\mu_1}$.
Then, no matter which output $d$ is chosen by~$P$ for $g=0$, the product $\mu_1\cdot\Pr\left[P\left(1,0\right)=d\right]$ will always be larger than $\left(1-\mu_1\right)\cdot\Pr\left[P\left(0,0\right)=d\right]$.
Consequently, the formula above is equal to $\mu_1$.
Thus, for highly uneven distributions of~$G$, 
the value of $V$ is independent of $P$ and, in particular, independent of the behavior of $P$ for $G=0$.
For example, in a setup where $\Pr\left[G=1\right]=0.98$, we could have an (unfair) program $P$ such that $\Pr\left[P\left(1,0\right)=1\right]\leq 1-(1-0.98)/0.98 \approx 0.979$ and $\Pr\left[P\left(0,0\right)=1\right]=0$, which would be indistinguishable from a (less unfair) program~$P'$ with $\Pr\left[P'\left(0,0\right)=1\right]>0$.
The reason for the indistinguishability is the high probability of $G=1$ which turns $G=1$ into the best guess for an attacker independently of $P$'s outcome.
Due to this, 
$P$'s decision for $G=0$ does not influence $V$.
\end{example}
This example demonstrates an extreme case of the above-mentioned problem that Conditional Vulnerability does not make a distinction as to why $G$ is easy to guess, which is obviously an an undesirable property for a fairness metric as it violates the second requirement from above.
This problem stems from the security context in which Conditional Vulnerability was originally defined:
When considering the amount of information leaked about a secret variable, it is a justified assumption that an attacker (whose likelihood of success we are trying to measure) will use public information on the probability distribution of the secret.
In this regard, however, the fairness setting differs from the security setting:
When measuring fairness, we consider information leakage of the group attribute problematic independently of the underlying group distribution.
For example, a biased credit algorithm would be deemed inacceptable independently of whether it discriminates against a group making up $10\%$ or $50\%$ of the population.
Therefore, we propose to measure the Conditional Vulnerability of a program under the assumption that $G$ is distributed uniformly.
If the measurement of $V$ is performed in a world with a uniform distribution of $G$, we can use it to compute a metric that we call the \emph{Fairness Spread} $S$ corresponding to a variant of demographic parity that is weighted by $U$.
In the following section, we are going to demonstrate that this metric satisfies the requirements from above.
Furthermore, we are going to prove that Fairness Spread is independent of $G$'s distribution which allows us to assume $G$'s uniformity when computing $S$ even for cases where $G$ is not in fact uniformly distributed.

\begin{table}[t]
     \caption{Results of quantitative analysis on examples in \Cref{lst:example_algos}: Each analysis took less than 3 seconds of computation time, count corresponds to $\left|\left\{
        \left(u,d\right) \in \mathcal{U}\times\mathcal{D} \mid
        \exists g \in \mathcal{G}. d=P\left(g,u\right)
    \right\}\right|$, $V$ and $S$ correspond to conditional vulnerability and fairness spread.}
    \label{tab:countersharp_experiment}   \centering
    \begin{tabular}{c|c|c|c|c}
         Program & $\left|\mathcal{U}\right|$ & Count & $V$ & $S$\\\hline
         \texttt{c1} & \phantom{0}10 & \phantom{0}20 & 0.2\phantom{0} & 1.0\\
         \texttt{c2} & \phantom{0}10 & \phantom{0}10 & 0.1\phantom{0} & 0.0\\
         \texttt{c3} & \phantom{0}10 & \phantom{0}12 & 0.12 & 0.2\\
         \texttt{c3-non-uniform}\footnotemark & 100 & 130 & 0.13 & 0.3\\
    \end{tabular}
\end{table}
\footnotetext{For the analysis of a non-uniformly distributed $U$ we resized $\mathcal{U}$ to 100 to more easily model the non-uniform distribution.}
\subsection{The Fairness Spread Metric}
To improve on the naive approach (see previous section), we propose an information-flow-based, quantitative fairness metric.
We call our metric Fairness Spread and initially introduce it for the case of independent $G$ and $U$ (this restriction is lifted in \Cref{sec:proxy}):
\begin{definition}[Fairness Spread]
    For independent variables $G \in \mathcal{G}$, $U \in \mathcal{U}$, $\left|\mathcal{D}\right|=2$ and a program $P$ we define the \emph{fairness spread} $S\left(G,U,P\right)$ as follows where ${\mu\left(g,u\right)=\Pr\left[P\left(G,U\right)=1 \mid G=g,U=u \right]}$:
    \begin{align*}
    \sum_{u \in \mathcal{U}}
    & \Pr\left[U=u\right] \cdot
    \max_{g_1,g_2\in\mathcal{G}} \left( \mu\left(g_1,u\right) - \mu\left(g_2,u\right) \right)
    \end{align*}
\end{definition}
In essence, Fairness Spread quantifies the weighted disparity in treatment between the most advantaged and most disadvantaged groups for each possible assignment of $U$ and is thus related to demographic parity. Fairness Spread is also related to a quantitative metric called Preference \cite{borca-tasciuc_provable_2022}, but differs from it by accounting for an individual's counterfactual outcome instead of comparing group conditional probability weight in absolute terms.
Note, that this metric satisfies our requirements %
from above: $S\left(G,U,P\right) = 0$  implies demographic parity, and if an algorithm further diverges from demographic parity for the most advantaged/disadvantaged group $S\left(G,U,P\right)$ increases.
Fairness Spread has an additional nice property, namely that it is independent of the distribution over~$G$:
\begin{lemma}[Independence of $G$]
\label{lemma:spread_g_independence}
    Let $G_1,G_2 \in \mathcal{G}$ be two random variables with independent arbitrary distributions with non-zero probabilities for all $g\in\mathcal{G}$.
    Then, for any program $P$ and any independent random variable $U \in \mathcal{U}$:
    $
    S\left(G_1,U,P\right) = S\left(G_2,U,P\right).
    $
\end{lemma}
This matches the intuition that discrimination of a group is undesirable irrespective of the group's population share.
By fixing $G$ to be uniformly distributed when computing $V\left(G \mid P\left(G,U\right),U\right)$ the obtained value can be mapped exactly to the notion of Fairness Spread as seen in \Cref{thm:vulnerability_fairness_spread}.
\begin{theoremE}[Conditional Vulnerability measures Fairness Spread][end,restate]
\label{thm:vulnerability_fairness_spread}
Let $G\in\mathcal{G}$ and $U\in\mathcal{U}$ be independent random variables and let $G$ be uniformly distributed. Then:
\[
S\left(G,U,P\right) \quad=\quad \left|\mathcal{G}\right| \cdot V\left(G \mid P\left(G,U\right),U\right) - 1
\]
\end{theoremE}
\begin{proofE}
For proof see \Cref{fig:proof_vulnerability_fairness_spread}
\begin{figure*}
    Let $n$ be the cardinality $\left|\mathcal{G}\right|$.
    Then:  
\begin{align*} & \left|\mathcal{G}\right|\cdot V\left(G \mid P\left(G,U\right),U\right) - 1\\
    &=n \sum_{u \in \mathcal{U}} \sum_{d \in \mathcal{D}} \max_{g \in \mathcal{G}}  \left(
        \Pr\left[G=g \mid P\left(G,U\right) = d,U=u\right]\Pr\left[P\left(G,U\right) = d,U=u\right]
    \right)-1\\  
    &=n \sum_{u \in \mathcal{U}}
    \Pr\left[U=u\right]
    \sum_{d \in \mathcal{D}} \max_{g \in \mathcal{G}}  \left(
        \Pr\left[G=g \mid P\left(G,U\right) = d,U=u\right]\Pr\left[P\left(G,U\right) = d \mid U=u\right]
    \right)-1\\
    &=n \sum_{u \in \mathcal{U}}
    \Pr\left[U=u\right]
    \sum_{d \in \mathcal{D}} \max_{g \in \mathcal{G}}  \left(
        \frac{\Pr\left[ P\left(G,U\right)=d, G=g, U=u \right]}
        {\Pr\left[U=u\right]}
    \right)-1\\
    &=n \sum_{u \in \mathcal{U}}
    \Pr\left[U=u\right]
    \sum_{d \in \mathcal{D}} \max_{g \in \mathcal{G}}  \left(
        \Pr\left[P\left(G,U\right)=d \mid G=g,U=u \right]
        \Pr\left[G=g\right]
    \right)-1\\
    &= \frac{n}{n}\sum_{u \in \mathcal{U}}
    \Pr\left[U=u\right]
    \sum_{d \in \mathcal{D}} \max_{g \in \mathcal{G}}  \left(
        \Pr\left[P\left(G,U\right)=d \mid G=g,U=u \right]
    \right)-1\\
    &= \sum_{u \in \mathcal{U}}
    \Pr\left[U=u\right]\left(
     \max_{g \in \mathcal{G}}  \left(
        \Pr\left[P\left(G,U\right)=1 \mid G=g,U=u \right]
    \right)
    +
    \max_{g \in \mathcal{G}}  \left(
        \Pr\left[P\left(G,U\right)=0 \mid G=g,U=u \right]
    \right)\right)-1\\
\intertext{-- using the abbreviation $\gamma_i = \Pr\left[P\left(G,U\right)=1 \mid G=i,U=u\right]$ for $i \in \left\{1,\dots,n\right\}$ --}
&= \sum_{u \in \mathcal{U}}
    \Pr\left[U=u\right]\left(
    \max\left(\gamma_1,\dots,\gamma_n\right) + 
    \max\left(\left(1-\gamma_1\right), \dots, \left(1-\gamma_n\right)\right)\right)
    -1\\
\intertext{-- assuming w.l.o.g.\ that the $\gamma_i$ are sorted, i.e., $\gamma_1 \leq \gamma_2 \leq \cdots \leq \gamma_n$ --}
&= \sum_{u \in \mathcal{U}}
    \Pr\left[U=u\right]\left(
    \gamma_n + 
    \left(1-\gamma_1\right)\right)
    -1\\
    &= \sum_{u \in \mathcal{U}}
    \Pr\left[U=u\right]\left(
     \Pr\left[P\left(G,U\right)=1 \mid G=n,U=u \right]
    -\Pr\left[P\left(G,U\right)=1 \mid G=1,U=u \right]\right)\\
    &= \sum_{u \in \mathcal{U}}
    \Pr\left[U=u\right]\left(
    \max_{g_1,g_2\in\mathcal{G}} \left(
\Pr\left[P\left(G,U\right)=1 \mid G=g_1,U=u \right]
    -\Pr\left[P\left(G,U\right)=1 \mid G=g_2,U=u \right]
    \right)
    \right)\\
    &= 
    S\left(G,U,P\right)
\end{align*}
\caption{Proof of \Cref{thm:vulnerability_fairness_spread}}
\label{fig:proof_vulnerability_fairness_spread}
\end{figure*}
\end{proofE}
Since we already discussed the suitability of Fairness Spread as a metric for fairness, 
\Cref{thm:vulnerability_fairness_spread} %
provides a sensible formalism to measure an algorithm's fairness.
Furthermore, \Cref{lemma:spread_g_independence} enables us to compute $S$ even in setting where $G$ is not uniformly distributed thus ensuring a wide applicability.

\subsection{Application to Examples}
We can now quantify how fair resp.\ unfair the programs from \Cref{lst:example_algos} are with respect to the metric of Fairness Spread.
To this end, we analyzed all three programs for a setting with 10 groups (e.g., age ranges) and a credit score of $U$ between 1 and~10.
Initially, we assumed a uniform distribution of credit scores. In addition, we analyzed \texttt{c3} (\Cref{lst:example_algos}~(b)) using a wrapper program that adjusted the probability of a credit score between 6 to 7 to be 30\% (instead of 20\% in the uniform case). %
All analyses are summarized in \Cref{tab:countersharp_experiment} and were computed using the counterSharp tool which allows the  quantitative analysis of C-programs.
Recent advances in (approximate) Model Counting ~\cite{DBLP:series/faia/ChakrabortyMV21,DBLP:conf/cav/YangM23} and quantitative program analysis show that this technique scales even to programs with more than 1000 LOC~\cite{teuber_quantifying_2021}.
The computed results match our previously developed intuition for the three programs: While \texttt{c1} is entirely unfair due to its decision-making solely based on \texttt{group}, the program \texttt{c3} is less fair than \texttt{c2}, but fairer than \texttt{c1}, due to its limited use of the \texttt{group} variable.
In the table's last line, we see that \texttt{c3} becomes more unfair according to $S$ if a larger share of the population have a credit score of 6 or~7.
That makes sense because in this case a larger share of individuals with \texttt{group >= 6} will be denied credit in comparison to younger individuals with the same score.

\section{Modeling Dependent Variables}
\label{sec:proxy}
The previous sections assume the independence of $G$ and~$U$, which is in general a strong -- if not too strong -- assumption for fairness analyses.
This raises the question of how our approach can be extended to the analysis of disparate impact in the presence of dependencies between $G$ and $U$.
Fortunately, as our approach analyzes programs, we can also analyze the structural equations of a causal model.
This observation leads to the following lemma for binary classifiers where $\hat{P}_C$ is the program composed of $P$ and a structural model $C$ as defined in \Cref{sec:background}:
\begin{lemmaE}[Coincidence of Fairness Spread and Counterfactual Fairness]
    \label{lem:coincidence_counterfactual}
    Let $P$ be some program with $\left|\mathcal{D}\right|=2$ and $C$ some causal model over background variables $B$. Then,
    $P$ is counterfactually fair \enspace iff\enspace $S\left(G,B,\hat{P}_C\right)=0$\enspace.
\end{lemmaE}
\begin{proofE}
    By denoting as $\left(g,u\right)=C\left(b\right)$ the statement that $C$ returns $G=g$ and $U=u$ for an input $b$ we can rewrite the counterfactual fairness property by requiring for all $g_1,g_2 \in \mathcal{G}, u \in \mathcal{U}$ that
    $\Pr\left[
    \hat{P}_C\left(g_i,B\right) = 1 \middle|
    \left(g_1,u\right) = C\left(B\right)
    \right]$ is equal for both $i\in\left\{1,2\right\}$.
    This is equivalent to the property that the $\Pr\left[B=b\right]$ weighted sum of 
    $\Pr\left[
    \hat{P}_C\left(g_i,b\right) = 1 \middle|
    \left(g_1,u\right) = C\left(b\right),B=b
    \right]$ over all $b$ is the same for both $i\in\left\{1,2\right\}$.
    Factoring out the now deterministic event $\left(g_1,u\right) = C\left(b\right)$ we can show that this is equivalent to the following property:
    \begin{align*}
    &0 = \sum_{b}
    \Pr\left[B=b\right] \cdot\\
    &\left|
    \Pr\left[
    \hat{P}_C\left(g_1,B\right) = 1 \middle|
    B=b
    \right] - 
    \Pr\left[
    \hat{P}_C\left(g_2,b\right) = 1 \middle|
    B=b
    \right]
    \right|
    \end{align*}
    The difference over the two probabilities is larger than $0$ and smaller than the maximum of this difference over all $g_i,g_j \in \mathcal{G}^2$.
    Thus, if $S\left(G,B,\hat{P}_C\right)=0$ then the absolute value is $0$ as well.
\end{proofE}
As a consequence of Lemma~\ref{lem:coincidence_counterfactual}, a binary decision procedure $P$ is counterfactually fair iff $\hat{P}_C\left(G,B\right)$ satisfies unconditional noninterference allowing the application of the entire machinery of \Cref{sec:qualitative-information-flow}.
We have furthermore once again satisfied the first of the two requirements for a fairness metric outlined in \Cref{sec:quantative-information-flow}, as a fairness spread of 0 coincides with counterfactual fairness.
Addressing the second requirement (a larger value implies more unfairness),
we can analyze the probability of a random individual having a counterfactual.
To this end, we define a difference function which, given a concrete  assignment to the background variables $B=b$ and a program $P$, provides the maximal deviation of counterfactuals. This can be defined as $\mathrm{Diff}_C\left(P,b\right)=\max_{g\in\mathcal{G}} \left|\hat{P}_C\left(b\right) - \hat{P}_C\left(g,b\right)\right|$. Note, that for binary decisions $\mathrm{Diff}_C$ is $0$ or $1$.
Using this function, we can compute the probability of a random individual having a deviating counterfactual, i.e., $\Pr\left[\mathrm{Diff}_C\left(P,B\right)=1\right]$ which we can bound through Fairness Spread:
\begin{lemmaE}
    \label{lem:prob_counterfactual_bound}
    Let $P$ be some program and $C$ some causal model over background variables $B$, and let $\left|\mathcal{D}\right|=2$. Then:
    $
    \Pr\left[\mathrm{Diff}_C\left(P,B\right)=1\right] \leq S\left(G,B,\hat{P}_C\right)
    $.
    And for $\left| \mathcal{G} \right| = 2$, the two terms are equal.
\end{lemmaE}
\begin{proofE}
We can rewrite $\Pr\left[\mathrm{Diff}_C\left(P,B\right)=1\right]$ as follows:
\begin{align*}
    &\sum_b \Pr\left[B=b\right] \cdot 
     \Pr\left[\mathrm{Diff}_C\left(P,B\right)=1 \middle| B=b\right]
\intertext{Since we know that $\mathrm{Diff}$ is 0 or 1 for binary decisions:}
    =\enspace &
    \sum_b \Pr\left[B=b\right] \cdot 
    \underbrace{\max_{g'\in\mathcal{G}} \left| \hat{P}_C\left(s\right) - \hat{P}_C\left(g',s\right) \right|}_{\theta}
\end{align*}
In any case we know that:
\[
\theta \leq \max_{g\in\mathcal{G}} \max_{g'\in\mathcal{G}} \left| \hat{P}_C\left(g,s\right) - \hat{P}_C\left(g',s\right) \right|
\]
Moreover, for $\left|\mathcal{G}\right|=2$ we know that the two terms are equal.
\end{proofE}
Note, that the probability term bounded by \Cref{lem:prob_counterfactual_bound} is -- just like $S$ -- independent of $G$'s distribution as it merely talks about the existence of a deviating counterfactual w.r.t.\ any other group.
As an example, consider the program in \Cref{lst:mixed}:
We previously established that the program is less fair than \Cref{lst:score_only} according to Fairness Spread.
We will now reevaluate the fairness of this program assuming a causal model.
To this end, we assume \texttt{score} is provided externally and computed based on knowledge about \texttt{income} and a client's \texttt{zipCode} (encoded as a ranking) which we cannot observe.
The causal model $C$ can be formalized as follows where we assume that different age groups live in different 
regions:\footnote{The causal models presented in this paper are just examples. We do not claim that the presented causalities do in fact exist.}
\begin{align*}
    \texttt{group} &\coloneqq\quad B_1\\
    \texttt{income} &\coloneqq\quad B_2\\
    \texttt{zipCode} &\coloneqq\quad  \textbf{if}\quad\left(\texttt{group}\geq 6\right)\quad B_3\quad\textbf{else}\quad B_4\\
    \texttt{score} &\coloneqq\quad \texttt{income} + \texttt{zipCode}
\end{align*}
We now encode the causal model as a program and analyze the programs \texttt{c2} and \texttt{c3} using the tooling described in \Cref{sec:quantative-information-flow}.\footnote{We assume ${B_1 \sim \mathcal{U}_d\left(0,9\right)}$, $ {B_2 \sim \mathcal{U}_d\left(0,9\right)}$, ${B_3 \sim \mathcal{U}_d\left(-1,5\right)}$, ${B_4 \sim \mathcal{U}_d\left(-3,3\right)}$ where $\mathcal{U}_d$ is the discrete uniform distribution.}
The analysis shows that you are more likely to have a deviating counterfactual when processed by \texttt{c2} than when processed by \texttt{c3}:
While we obtain a Fairness Spread of $0.23$ for \texttt{c3}, the Fairness Spread of \texttt{c2} is $0.27$ (analyses were performed in less than 2s using counterSharp).
This result remains the same when, only considering the groups $\texttt{group}=5$ and $\texttt{group}=6$.
Intuitively, this can be understood as follows:
If your age attribute is $5$ and your data is processed by \texttt{c2} you have a 27\% chance that you have a counterfactual with a deviating result for $\texttt{group}=6$.
On the other hand, when your data is processed by \texttt{c3} this probability drops to~23\%, indicating that \texttt{c3} is fairer than \texttt{c2} w.r.t.\ the above causal model.
This is due to the (unfair) influence of \texttt{group} on \texttt{score} which \texttt{c2}  (partially) counteracts.
This demonstrates that analyses assuming independence are not always sufficient and can even lead to misleading results (indicated by the difference to the results in \Cref{tab:countersharp_experiment}, where \texttt{c2} is fairer than \texttt{c3}).

Through the strong relation of Conditional Vulnerability and Counterfactual Fairness expressed in \Cref{lem:coincidence_counterfactual,lem:prob_counterfactual_bound} we can seamlessly apply qualitative and quantitative information flow analyses in contexts with dependent variables by suitably encoding their corresponding causal model.

While in the example above the disparate treatment based on \texttt{zipCode} leads to a disparate impact which is deemed inacceptable, there may be cases in where such disparities are justified.
In such cases, an extension called path-dependent counterfactual fairness~\cite{kusner2017counterfactual,loftus_causal_algorithmic_fairness} can be used which is equally supported by our Information-Flow based approach (see \Cref{apx:path_specific}).

\section{Case Studies}
\label{sec:examples}
We applied our approach to two case-studies, showcasing both the qualitative (\Cref{subsec:case_study:tax}) as well as the quantitative (\Cref{subsec:case_study:car_insurance}) approach\footnote{Artifact available at~
\cite{Artefact}
}:
First, we show how our approach can be used to analyze fairness in the context of German Tax computation.
Moreover, we demonstrate our causal quantitative analysis on the computation of car insurance premiums. %
More applications can be found in \Cref{apx:further_examples}.

\subsection{Analysis of German Tax Computation}
\label{subsec:case_study:tax}
German employers are required by law to deduct an employee's wage tax from their salary before payout.
To this end, it is necessary that employers can reliably compute an employee's wage tax.
Errors in this computation could lead to payouts of parts of the wage being delayed for over a year (in case the employee files a tax statement) or never being paid out.
For that purpose, the German government provides a PDF file containing 26 pages of flowcharts outlining the computation of the german wage tax~\cite{Bundesministerium_PAP} as well as a Java implementation~\cite{Bundesministerium_PAP_Java}.
This software consists of 1.5k LOC and solely operates on a global state while reusing many intermediate variables.
As Germany has a church tax, which this Java program also computes, it requires a person's religious affiliation as input (passed as an integer). This raises the question of whether (e.g., through a program error) a person's religious affiliation also influences the wage tax (rather than only influencing the church tax).
Notably, this property cannot be checked through exhaustive testing as the program has 35 inputs (21 of which can be arbitrary integers).
Therefore, we analyzed the Java programs from the years 2015--2023 using the information flow analysis tool Joana~\cite{tooljoana2013atps,joana14it} and were able to show that there is no disparate treatment in the wage tax computation.
Note, that this analysis was fully automated and demonstrates the scalability of this approach.

\subsection{Analysis for Car Insurance Decisions}
\label{subsec:case_study:car_insurance}
This case study is adapted from \cite{kusner2017counterfactual}, who analyzed the example of a car insurance company increasing premiums if an individual registers a red car.
Instead of color we consider engine power, which in our causal model is influenced both by \texttt{gender} and \texttt{aggressive}ness of the driver.
The probability of an accident is solely determined by aggressiveness.
The causal model can be summarized as follows where the insurance company wants to increase the premium whenever $\texttt{accident}>12$:
\begin{align*}
    \texttt{gender} &\coloneqq B_1\\
    \texttt{aggressive} &\coloneqq B_2\\
    \texttt{engine} &\coloneqq 3*\texttt{gender} + 8*\texttt{aggressive}\\
    \texttt{accident} &\coloneqq 20*\texttt{aggressive}
\end{align*}
Assuming that \texttt{aggressive} cannot be observed directly, the insurance company has the choice of either building a linear classifier only considering the observable variable \texttt{engine} or considering both \texttt{engine} and \texttt{gender} (see \Cref{apx:car_insurance}).
We quantitatively analyzed the two programs implementing the two possibilities w.r.t.\ the above causal model and find that the algorithm that takes both \texttt{engine} and \texttt{gender} into account is counterfactually fair %
while the classifier only considering \texttt{engine} has a Fairness Spread of $0.36$.\footnotemark~ 
This is, because considering \texttt{gender} removes the aggressiveness independent imbalance introduced through engine power.
Note, that the latter algorithm does not process the group variable. But, using causal modeling and information flow, we can still find analyze its impact.
\footnotetext{This assumes $B_1 \sim \mathcal{U}_d\left(0,1\right),B_2 \sim \mathcal{U}\left(0,1\right)$}

\section{Discussion and Conclusion}
This paper demonstrates numerous interconnections between the fields of Information Flow and Algorithmic Fairness.
In particular, our work demonstrates that pre-existing Information Flow approaches can be used to analyze fairness questions.
We present an approach for the analysis of concrete program code instead of abstract models or empirical evaluations -- even taking into account variable dependencies through the use of causal modeling.
We believe that the ability to analyze a decision procedure's actual code is one of the major strengths of the approach presented in this paper.
Some potential limitations and ethical considerations of our approach are discussed in \Cref{apx:ethics}
In future work, we plan to expand the tooling available for the analysis of information flow to ML systems allowing for the analysis of such decision making systems in a manner similar to classical program code.

\ifAAAI
\appendix
\refstepcounter{section}
\refstepcounter{subsection}
\label[appendix]{apx:conditional-information-flow}
\refstepcounter{subsection}
\label[appendix]{apx:compute_cond_vuln}
\refstepcounter{subsection}
\label[appendix]{apx:path_specific}
\refstepcounter{subsection}
\label[appendix]{apx:car_insurance}
\refstepcounter{subsection}
\label[appendix]{apx:further_examples}
\fi

\section*{Ethical Statement}
\label{apx:ethics}
The techniques outlined in this work allow for the analysis of complex, intricate fairness properties on the program code level. This approach has multiple advantages:
\begin{inparaenum}
    \item By starting from a formal fairness specification, the fairness properties an algorithm is supposed to possess are made explicit and auditable by computer scientists as well as experts from other fields. This is particularly beneficial if justified exceptions lead to conditional or restricted fairness as the exceptions are made explicit.
    \item By providing means to formally prove that a program conforms to a given formal fairness specification, this approach can provide a guarantee which goes beyond the guarantees attainable through purely empirical or statistical analysis;
    \item Through its distinction between disparate treatment and impact, information-flow analysis can help in uncovering and mitigating biases in an algorithm's code.
\end{inparaenum}

Nonetheless, employing this technique brings certain hazards.
In particular, a software engineer may be tempted to use off-the-shelf information flow-analysis tools to put a ``seal'' on their software without fully understanding or publicizing the assumptions used in a proof of noninterference, i.e., what specification the program is actually proven to satisfy.
Therefore, we recommend that an information-flow-based assessment of a software's fairness should be publicized in combination with all assumptions for and limitations to this fairness guarantee.
This makes such an assessment auditable and ensures that third parties can check whether assumptions hold (e.g., that the conditions of conditional fairness are actually justified) or critique potentially misguided use of program-analysis-based fairness guarantees.

\section*{Acknowledgements}
This work was supported by funding from the pilot program Core-Informatics of the Helmholtz Association (HGF).
It was furthermore supported by KASTEL Security Research Labs.
We thank Dr. Alexander Weigl and Dr. Michael Kirsten for helpful discussions and the anonymous reviewers for their  feedback on earlier drafts of this paper.

\bibliography{main}

\clearpage
\onecolumn
\input{corrigendum}

\twocolumn

\ifAAAI
\else
\clearpage
\appendix
\section{Supplementary Material}
\subsection{Conditional Information Flow and Fairness}
\label{apx:conditional-information-flow}
To demonstrate conditional information flow, we will once again consider the example presented in \Cref{lst:mixed}:
In terms of information flow, the program would have permission to declassify the variable \texttt{group} iff the following condition is satisfied:
\[
\psi_{\text{declass}} \enspace\equiv\enspace 6 \leq u \land u < 8 \land g \geq 6.
\]
If the group attribute is only declassified in this legally justified exception, we should still be able to prove noninterference for all other cases, i.e., whenever the values of $g$ and $u$ satisfy $\neg \psi_{\text{declass}}$ (denoted as $\left(g,u\right) \vDash \neg \psi_{\text{declass}}$).
To this end, we introduce Conditional Information Flow:
\begin{definition}[Conditional Information Flow]
Let $\psi$ be some constraint on $g$ and $u$.
A program $P$ satisfies \emph{Conditional Information Flow} for $\psi$ iff, for all $g_1,g_2 \in \mathcal{G}$ and all $u \in \mathcal{U}$,
\[\left(g_1,u\right) \vDash \neg\psi \mbox{\ and\ } \left(g_2,u\right) \vDash \neg\psi ~\mbox{implies}~ P\left(g_1,u\right) = P\left(g_2,u\right).
\]
\end{definition}
We can once again relate this notion to the fairness literature as conditional information flow corresponds to the concept of Conditional Demographic Parity~\cite{mitchell2018predictionbased}:
\begin{lemmaE}[Conditional Information Flow implies Conditional Demographic Parity]
    \label{lem:conditional_dem_parity}
    Let $P$ be a program and let $G,U$ be distributed arbitrarily, but independently.
    If $P$ satisfies Conditional Information Flow for $\psi$, then the outcome of~$P$ satisfies \emph{conditional demographic parity} for $\neg\psi$, i.e., for all $d \in \mathcal{D}, g_1,g_2 \in \mathcal{G}$:
    \[
    \Pr\left[P\left(g_1,U\right)=d \mid \neg\psi \right] \enspace=\enspace \Pr\left[P\left(g_2,U\right)=d \mid \neg\psi \right]
    \]
\end{lemmaE}
\begin{proofE}
    The proof is analogous to the proof in \Cref{lem:uncond_dem_parity}, except that we restrict the domain to only consider those $\left(g_1,u\right),\left(g_2,u\right)$ that satisfy~$\psi$.
\end{proofE}
For the case of \Cref{lst:mixed}, Conditional Information Flow holds for the above condition~$\psi_{\text{declass}}$ since there is only a disparity in results when $\psi_{\text{declass}}$ is satisfied.
Again, it is possible to formally prove this using the \KeY tool.

We can compare this to the case where line 2 in \Cref{lst:mixed} checks \texttt{group >= 8} instead of \texttt{group >= 6}:
While this program would still satisfy Conditional Information Flow with respect to $\psi_{\text{declass}}$, it would not satisfy Restricted Information Flow with respect to $R$ as seen in \Cref{sec:restricted-information-flow} as we discriminate against a subset of a class defined by $R$.
As demonstrated by the example, Restricted Information Flow often provides a more fine-grained view on the Fairness of an algorithm.

While this approach ensures that all individuals satisfying $\neg\psi_{\text{declass}}$ are treated equally, the algorithm is still able to use the information of whether $\neg\psi_{\text{declass}}$ is satisfied for a given individual in order to compute its outcome.
Thus, in this setting, the algorithm will always be allowed to use at least this one bit of information on protected group attributes for its computation.

\subsection{Computation of Fairness Spread}
\label{apx:compute_cond_vuln}
Conditional Vulnerability is not only a suitable metric but can furthermore be easily computed for uniform input distributions of $U$ with the help of preexisting information-flow analysis techniques making it computable using off-the-shelf tools.
Moreover, we can even allow for more complicated distributions for $U$ within our analyzed program, thus lifting the limitation to $U$'s distribution.
Using the results by \cite{Smith2009}, we can derive a formula for computing conditional vulnerability in the presence of unprotected variables:
\begin{lemmaE}[Computation of $V$ through counting][end,restate]
\label{lemma:compute_vulnerability}
    Let $G$ and $U$ be uniformly distributed and $P$ be deterministic.
 Then we can compute $V\left(G \mid P\left(G,U\right),U\right)$ through the following formula:
    \[
    \quad \frac{\left|\left\{
        \left(u,d\right) \in \mathcal{U}\times\mathcal{D} \mid
        \exists g \in \mathcal{G}. d=P\left(g,u\right)
    \right\}\right|}{\left|\mathcal{U}\right|\left|\mathcal{G}\right|}
    \]
\end{lemmaE}
\begin{proofE}
Our proof makes use of a result by \cite{Smith2009}:
Let $G$ be distributed uniformly. For a fixed $U=u\in\mathcal{U}$,  the Conditional Vulnerability $V\left(G \mid P\left(G,U\right),U=u\right)$ of a deterministic program $P$ can be computed as follows:
\[
 \frac{\left|\left\{d \in \mathcal{D} \mid \exists g \in \mathcal{G}. P\left(g,u\right)=d\right\}\right|}{\left|\mathcal{G}\right|}
\]
Using this result we can rewrite $V\left(G \mid P,U\right)$ as follows:
\begin{align*}
    &\sum_{u \in \mathcal{U}} \Pr\left[U=u\right] V\left(G \mid P,U=u\right)\\
    =\enspace&\frac{1}{\left|\mathcal{U}\right|\left|\mathcal{G}\right|}
    \sum_{u \in \mathcal{U}} \left| \left\{d \in \mathcal{D} \mid \exists g \in \mathcal{G}. P\left(g,u\right)=d\right\} \right|\\
    =\enspace&\frac{\left|\left\{
        \left(u,d\right) \in \mathcal{U}\times\mathcal{D} \mid
        \exists g \in \mathcal{G}. d=P\left(g,u\right)
    \right\}\right|}{\left|\mathcal{U}\right|\left|\mathcal{G}\right|}
\end{align*}
\end{proofE}
This manner of computation also provides a good intuition for the exact meaning of Conditional Vulnerability:
By counting the number of possible outcomes $d$ for a given unprotected value $u$, we essentially count the number of cases where the same unprotected input can yield different decisions.
Since $P$ is deterministic, a differing decision for a given $d$ implies that $P$ makes a decision based on the protected group attribute~$g$.
In practice, we can compute the Conditional Vulnerability of a program using propositional model counting techniques.
Through \Cref{lemma:spread_g_independence}, the result from \Cref{lemma:compute_vulnerability} not only provides us with a method for computing $S\left(G,U,P\right)$ for uniform distributions over $G$, but for arbitrary distributions over $G$.
Furthermore, the restriction of $U$ to uniform distributions is not relevant for the practical application of the approach, as we can wrap $P$ into a larger program that constructs more complicated distributions for the unprotected input of $P$ from a uniformly distributed $U$.

\subsection{Analysis of Path-specific Counterfactual Fairness}
\label{apx:path_specific}
The example in \Cref{sec:proxy} implicitly assumes that disparate treatment based on \texttt{zipCode} is unjustified if it leads to disparate impact with respect to the \texttt{group} attribute.
This may not always be the case: For example, in cases where the client's property serves as a collateral for a loan the property value (partially determined by it's location and thus its zip code) may be of relevance.
To this end, it may be necessary to analyze an algorithm w.r.t. path-dependent counterfactual fairness~\cite{kusner2017counterfactual,loftus_causal_algorithmic_fairness} which is also supported by our approach through the introduction of additional unprotected variables.
To this end, assume a set of indices $P$ such that it is considered fair to use a causal model's variables $V_i$ for $i \in P$ in order to compute a decision even in the case where there exists a causality $G \rightarrow V_i$.
A classic example that can be analyzed through the lense of path-specific counterfactual fairness are the gender bias claims in UC Berkley's admissin process in the 1970s~\cite{kusner2017counterfactual} as course selection (despite gender disparities) is often judged as a free choice of the applicants.

In order to analyze path-specific counterfactual fairness of $P$ w.r.t. the causal model $C$ we will once again analyze $\hat{P}_C\left(B\right)$ and $\hat{P}_C\left(G,B\right)$ where $B$ is the (public) assignment of background variables and $G$ is the protected (secret) counterfactual group assignment.
However, we are going to extend the program by an additional public input variable $V_P$ which assigns the path specific variables.
By assigning $V_P$ a public status, we assert that its value must be equal when comparing the outcomes for two different groups $g_1,g_2$.
Given a concrete value $B=b$ we can compute the factual group $g$ based on $b$.
The domain of $V_P$ is then any assignment of the variables $V_P$ which is in accordance with the $b$ and $g$ according to the causal model $C$.
This can easily be encoded as a requirement/assumption when evaluating the qualitative/quantitative information flow.

We now demonstrate this on a concrete example:
Reconsidering the causal model of \Cref{sec:proxy}, we can consider the case where disparate treatment based on \texttt{zipCode} is considered fair (e.g. due to its relevance for the decision or due to a judgement that one is free to choose their place of residence).
In this case, we extend $\hat{P}_C$ by a new public input variable \texttt{zipCode} which determines the considered zip code path, i.e. it assigns \texttt{zipCode} to a fixed value across counterfactual evaluations w.r.t. different groups.
In this case (reevaluating in the same setting as in \Cref{sec:proxy}), \texttt{c2} (\Cref{lst:score_only}) is once again counterfactually fair (i.e. its Fairness Spread is $0$) while \texttt{c3} (\Cref{lst:mixed}) has a Fairness Spread of $0.19$ meaning that $19\%$ of individuals have a path-specific deviating counterfactual.
The reduction of Fairness Spread for \texttt{c3} w.r.t. the not path-specific analysis is due to the fact that some cases which could previously appear are no longer possible now:
Previously an individual with $\texttt{income}=3$ and $\texttt{zipCode}=2$ could have had a deviating (older) counterfactual with $\texttt{zipCode}=5$ which would have gotten a positive result for their loan application.
This is no longer possible now as the counterfactual is required to have the same \texttt{zipCode} whose range in the analysis is furthermore dependent on $B_1$.

\subsection{Car Insurance Classifiers}
\label{apx:car_insurance}
The two linear classifiers for the computation of car insurance premiums can be found in \Cref{lst:car_insurance}: The two classifiers were constructed using linear regression on data sampled from the causal model in \Cref{subsec:case_study:car_insurance}.
The insurance's decision criterion is whether the predicted value of \texttt{accident} is above 12.
As the integer inputs represent fractional numbers (the number 1 is represented as 100), the coefficients in \Cref{lst:car_insurance} are larger.

\begin{figure}[t]
\centering
\begin{lstlisting}[language=Python,style=mystyle,escapechar=|]
def insurance_premium(engine):
    return (188*engine_power >= 1049);
def insurance_premium_fair(gender, engine):
    return (-749*race + 248*engine  >= 1121);
\end{lstlisting}
\caption{Two algorithms for computing an additional car insurance premium: One algorithm only considers engine power while the other algorithm considers engine power and gender.}
\label{lst:car_insurance}
\end{figure}

\subsection{Further Applications}
\label{apx:further_examples}

\begin{figure}
\begin{subfigure}[t]{0.5\linewidth}
\vskip 0pt
\begin{tabular}{c}
\begin{lstlisting}[language=Python,style=mystyle,escapechar=|]
def credit(gender, amount):
  if gender==0:
    return (amount <= T)
  else:
    return (amount > (10-T))
\end{lstlisting}
\end{tabular}
\subcaption{Algorithm.}
\label{lst:unfair_credit}
\end{subfigure}\hspace{0.05\linewidth}
\begin{subfigure}[t]{0.44\linewidth}
\vskip -1em
\begin{tikzpicture}
\begin{axis}[height=2.5cm,width=3.8cm,
    x label style={at={(axis description cs:0.5,-0.3)},anchor=north},
    y label style={at={(axis description cs:-0.3,.5)},anchor=south},
   xlabel=T,
   ylabel={$S\left(G,U,P\right)$}]
 \addplot table [x=T, y=S]{unfair_credit.dat};
\end{axis}
\end{tikzpicture}
\subcaption{Fairness Spread for different values of \texttt{T} in \Cref{lst:unfair_credit}}
\label{fig:unfair_spread}
\end{subfigure}
\caption{Unfair credit algorithm}
\end{figure}
The previous sections already presented some example analyses of simple credit decision algorithms.
As another exemplary application in the loan allocation setting, we consider an algorithm that satisfies demographic parity but does not conform to unconditional information flow. The program can be found in \Cref{lst:unfair_credit}:
For $\texttt{amount}\in\left[1,10\right]$ uniformly distributed and $\texttt{T}=5$, the binary outcome of \texttt{credit} would conform to the definition of demographic parity as the group-conditional probability of obtaining a credit is equal.
Nonetheless, this program would not satisfy unconditional noninterference as the outcomes for a fixed unprotected assignment of \texttt{amount} differ based on the value of \texttt{gender}.
This is an example of a program where there are good reasons to consider it unfair due to the use of the variable \texttt{gender} in it's decision process despite a demographic parity guarantee, because the algorithm only awards small credits for \texttt{gender==0} while only awarding large credits otherwise.
This shows that the Information Flow based fairness notions we have introduced in \Cref{sec:qualitative-information-flow} are stronger than demographic parity and can detect fairness deficiencies invisible to a naive statistical analysis of the binary outcome.
We can analyze this algorithm quantitatively for different values of $T$.
To this end, consider the plot in \Cref{fig:unfair_spread} which displays the Fairness Spread of the program for different threshold values \texttt{T}:
While the Fairness Spread is highest for $\texttt{T}=5$, it becomes less if \texttt{T} is increased or decreased which reduces the number of cases where the outcome differs based on gender.
The extreme cases where \texttt{T} is 0 or 10 then satisfy demographic parity where no one resp.\ everyone is awarded a credit.

As a final exemplary application, we consider an algorithm with more than two outcomes where notions of Restricted Information Flow can be applied.
To this end, we examine the case of algorithm based salary computation.
In the case of more than two outcomes, the results of \Cref{sec:qualitative-information-flow} are equally applicable.
For example, it may be the case that an employer provides a fixed social benefit for employees with disabilities, an ongoing pregnancy or a recent hospitalisation.
Each of these attributes could be considered a protected attribute.
In that case, the salary computation ought to satisfy restricted information flow with respect to $R\left(\texttt{pregnancy},\texttt{disability},\texttt{hospital}\right) = \texttt{pregnancy} \lor \texttt{disability}\lor\texttt{hospital}$.
Importantly, the algorithm must not distinguish between the three cases if the social benefit is set to the same  fixed value for all three cases.
If the benefit is cumulative, a program must satisfy restricted information flow with respect to a restricted classification $R$ which \textit{counts} the number of satisfied attributes (i.e., $R\left(\texttt{true},\texttt{false},\texttt{false}\right)=1$ and $R\left(\texttt{true},\texttt{true},\texttt{false}\right)=2$ etc.) in this case there are 4 classes while there exist only two classes for non-cumulative benefits.

\subsection{Proofs}
\printProofs

\fi
\end{document}

%% file: corrigendum.tex
\section*{Corrigendum}
\label{corrigendum}
\Cref{lem:restricted_information_flow} states that restricted information flow implies conditional demographic parity. This statement is wrong.
This was brought to our attention by Marco Favier, who also provided a concrete counterexample to the Lemma:

\paragraph{Counterexample.}
Assume $\mathcal{G}=\left\{A,B\right\}$ and $\mathcal{U}=\left\{\texttt{true},\texttt{false}\right\}$ with independent, uniform probabilities.
In this instance, we could define a restricted classification $R$ with $R\left(A,\texttt{true}\right)=R\left(B,\texttt{false}\right)=0$ and $R\left(A,\texttt{false}\right)=R\left(B,\texttt{true}\right)=1$.
Based on this setup, consider the following program:
\vspace*{0.25cm}
\begin{lstlisting}
if (R(G,U)==0)
    return U
else
    return !U
\end{lstlisting}
\vspace*{0.25cm}
This program satisfies the restricted information flow property with respect to the restricted classification given above; however, it does \emph{not} satisfy conditional demographic parity.
For example, under the constraint $R(G,U)=0$ all individuals with $G=A$ would be accepted while all individuals with $G=B$ would be rejected.
Consequently, Lemma 2 does not hold, and its proof is flawed because it ignores the change in distribution that stems from conditioning on, e.g., $R(G,U)=0$.

\paragraph{Wider Impact.}
Unfortunately, as a consequence of this counterexample, restricted information flow is not a suitable methodology to guarantee the (conditional) demographic parity of a given program.
Consequently, Section 3.2, Appendix A.1, and the final paragraph of Appendix A.5 of this paper must be disregarded as invalid.

It is worth noting that this flaw has no wider impact on the paper's other results:
The results on unconditional information flow and its relation to demographic parity (Section 3.1), as well as the results of Sections 4-7, are not affected by the invalid Lemma.
In particular, all results on quantitative measurements and their relationship to causal graphs are not affected by this flaw.

%% file: main.bbl
\begin{thebibliography}{36}
\providecommand{\natexlab}[1]{#1}

\bibitem[{Ahrendt et~al.(2005)Ahrendt, Baar, Beckert, Bubel, Giese,
  H{\"{a}}hnle, Menzel, Mostowski, Roth, Schlager, and
  Schmitt}]{ahrendt2005key}
Ahrendt, W.; Baar, T.; Beckert, B.; Bubel, R.; Giese, M.; H{\"{a}}hnle, R.;
  Menzel, W.; Mostowski, W.; Roth, A.; Schlager, S.; and Schmitt, P.~H. 2005.
\newblock The KeY tool.
\newblock \emph{Softw. Syst. Model.}, 4(1): 32--54.

\bibitem[{Ahrendt et~al.(2016)Ahrendt, Beckert, Bubel, H{\"{a}}hnle, Schmitt,
  and Ulbrich}]{KeYBook}
Ahrendt, W.; Beckert, B.; Bubel, R.; H{\"{a}}hnle, R.; Schmitt, P.~H.; and
  Ulbrich, M., eds. 2016.
\newblock \emph{Deductive Software Verification - The KeY Book - From Theory to
  Practice}, volume 10001 of \emph{LNCS}.
\newblock Cham: Springer.
\newblock ISBN 978-3-319-49811-9.

\bibitem[{Albarghouthi et~al.(2017)Albarghouthi, D'Antoni, Drews, and
  Nori}]{Albarghouthi2017}
Albarghouthi, A.; D'Antoni, L.; Drews, S.; and Nori, A.~V. 2017.
\newblock FairSquare: probabilistic verification of program fairness.
\newblock \emph{Proc. {ACM} Program. Lang.}, 1({OOPSLA}): 80:1--80:30.

\bibitem[{Bastani, Zhang, and Solar{-}Lezama(2019)}]{bastani2019probabilistic}
Bastani, O.; Zhang, X.; and Solar{-}Lezama, A. 2019.
\newblock Probabilistic verification of fairness properties via concentration.
\newblock \emph{Proc. {ACM} Program. Lang.}, 3({OOPSLA}): 118:1--118:27.

\bibitem[{Beckert et~al.(2013)Beckert, Bruns, Klebanov, Scheben, Schmitt, and
  Ulbrich}]{beckert_information_2013}
Beckert, B.; Bruns, D.; Klebanov, V.; Scheben, C.; Schmitt, P.~H.; and Ulbrich,
  M. 2013.
\newblock Information Flow in Object-Oriented Software.
\newblock In Gupta, G.; and Pe{\~{n}}a, R., eds., \emph{Logic-Based Program
  Synthesis and Transformation, 23rd International Symposium, {LOPSTR} 2013,
  Madrid, Spain, September 18-19, 2013, Revised Selected Papers}, volume 8901
  of \emph{LNCS}, 19--37. Cham: Springer.

\bibitem[{Beckert, Kirsten, and Schefczyk(2022)}]{beckert_algorithmic_2022}
Beckert, B.; Kirsten, M.; and Schefczyk, M. 2022.
\newblock Algorithmic {Fairness} and {Secure} {Information} {Flow} ({Extended}
  {Abstract}).
\newblock In Heitz, C.; Hertweck, C.; Viganò, E.; and Loi, M., eds.,
  \emph{European {Workshop} on {Algorithmic} {Fairness} ({EWAF} '22),
  {Lightning} round track}.

\bibitem[{Beutel et~al.(2019)Beutel, Chen, Doshi, Qian, Woodruff, Luu,
  Kreitmann, Bischof, and Chi}]{Beutel2019}
Beutel, A.; Chen, J.; Doshi, T.; Qian, H.; Woodruff, A.; Luu, C.; Kreitmann,
  P.; Bischof, J.; and Chi, E.~H. 2019.
\newblock Putting Fairness Principles into Practice: Challenges, Metrics, and
  Improvements.
\newblock In Conitzer, V.; Hadfield, G.~K.; and Vallor, S., eds.,
  \emph{Proceedings of the 2019 {AAAI/ACM} Conference on AI, Ethics, and
  Society, {AIES} 2019, Honolulu, HI, USA, January 27-28, 2019}, 453--459. New
  York, NY, USA: Association for Computing Machinery.

\bibitem[{Biondi et~al.(2018)Biondi, Enescu, Heuser, Legay, Meel, and
  Quilbeuf}]{biondi_scalable_2018}
Biondi, F.; Enescu, M.~A.; Heuser, A.; Legay, A.; Meel, K.~S.; and Quilbeuf, J.
  2018.
\newblock Scalable Approximation of Quantitative Information Flow in Programs.
\newblock In Dillig, I.; and Palsberg, J., eds., \emph{Verification, Model
  Checking, and Abstract Interpretation - 19th International Conference,
  {VMCAI} 2018, Los Angeles, CA, USA, January 7-9, 2018, Proceedings}, volume
  10747 of \emph{LNCS}, 71--93. Cham: Springer.

\bibitem[{Borca{-}Tasciuc et~al.(2023)Borca{-}Tasciuc, Guo, Bak, and
  Skiena}]{borca-tasciuc_provable_2022}
Borca{-}Tasciuc, G.; Guo, X.; Bak, S.; and Skiena, S. 2023.
\newblock Provable Fairness for Neural Network Models using Formal
  Verification.
\newblock In Alvarez, J.~M.; Fabris, A.; Heitz, C.; Hertweck, C.; Loi, M.; and
  Zehlike, M., eds., \emph{Proceedings of the 2nd European Workshop on
  Algorithmic Fairness, Winterthur, Switzerland, June 7th to 9th, 2023}, volume
  3442 of \emph{{CEUR} Workshop Proceedings}. CEUR-WS.org.

\bibitem[{Chakraborty, Meel, and
  Vardi(2021)}]{DBLP:series/faia/ChakrabortyMV21}
Chakraborty, S.; Meel, K.~S.; and Vardi, M.~Y. 2021.
\newblock Approximate Model Counting.
\newblock In Biere, A.; Heule, M.; van Maaren, H.; and Walsh, T., eds.,
  \emph{Handbook of Satisfiability - Second Edition}, volume 336 of
  \emph{Frontiers in Artificial Intelligence and Applications}, 1015--1045.
  {IOS} Press.

\bibitem[{Darvas, H{\"{a}}hnle, and Sands(2005)}]{DarvasHS05}
Darvas, {\'{A}}.; H{\"{a}}hnle, R.; and Sands, D. 2005.
\newblock A Theorem Proving Approach to Analysis of Secure Information Flow.
\newblock In Hutter, D.; and Ullmann, M., eds., \emph{Security in Pervasive
  Computing, Second International Conference, {SPC} 2005, Boppard, Germany,
  April 6-8, 2005, Proceedings}, volume 3450 of \emph{LNCS}, 193--209. Cham:
  Springer.

\bibitem[{Denning(1976)}]{Denning76}
Denning, D.~E. 1976.
\newblock A Lattice Model of Secure Information Flow.
\newblock \emph{Commun. {ACM}}, 19(5): 236--243.

\bibitem[{Denning(1975)}]{denning1975secure}
Denning, D. E.~R. 1975.
\newblock \emph{Secure information flow in computer systems.}
\newblock Ph.D. thesis, Purdue University.

\bibitem[{Dwork et~al.(2012)Dwork, Hardt, Pitassi, Reingold, and
  Zemel}]{DworkHPRZ12}
Dwork, C.; Hardt, M.; Pitassi, T.; Reingold, O.; and Zemel, R.~S. 2012.
\newblock Fairness through awareness.
\newblock In Goldwasser, S., ed., \emph{Innovations in Theoretical Computer
  Science 2012, Cambridge, MA, USA, January 8-10, 2012}, 214--226. New York,
  NY, USA: Association for Computing Machinery.

\bibitem[{{Federal Ministry of
  Finance}(2022{\natexlab{a}})}]{Bundesministerium_PAP_Java}
{Federal Ministry of Finance}. 2022{\natexlab{a}}.
\newblock Lohn- und Einkommenssteuerrechner: Service fuer Entwickler.

\bibitem[{{Federal Ministry of
  Finance}(2022{\natexlab{b}})}]{Bundesministerium_PAP}
{Federal Ministry of Finance}. 2022{\natexlab{b}}.
\newblock Programmablaufpläne für den Lohnsteuerabzug 2023.

\bibitem[{Ghosh, Basu, and Meel(2021)}]{Ghosh2021a}
Ghosh, B.; Basu, D.; and Meel, K.~S. 2021.
\newblock Justicia: {A} Stochastic {SAT} Approach to Formally Verify Fairness.
\newblock In \emph{Thirty-Fifth {AAAI} Conference on Artificial Intelligence,
  {AAAI} 2021, Thirty-Third Conference on Innovative Applications of Artificial
  Intelligence, {IAAI} 2021, The Eleventh Symposium on Educational Advances in
  Artificial Intelligence, {EAAI} 2021, Virtual Event, February 2-9, 2021},
  7554--7563. {AAAI} Press.

\bibitem[{Ghosh, Basu, and Meel(2022)}]{Ghosh_Basu_Meel_2022}
Ghosh, B.; Basu, D.; and Meel, K.~S. 2022.
\newblock Algorithmic Fairness Verification with Graphical Models.
\newblock \emph{Proceedings of the AAAI Conference on Artificial Intelligence},
  36(9): 9539--9548.

\bibitem[{Graf, Hecker, and Mohr(2013)}]{tooljoana2013atps}
Graf, J.; Hecker, M.; and Mohr, M. 2013.
\newblock Using JOANA for Information Flow Control in Java Programs - A
  Practical Guide.
\newblock In \emph{Proceedings of the 6th Working Conference on Programming
  Languages (ATPS'13)}, Lecture Notes in Informatics (LNI) 215, 123--138.
  Springer Berlin / Heidelberg.

\bibitem[{Grgic-Hlaca et~al.(2016)Grgic-Hlaca, Zafar, Gummadi, and
  Weller}]{grgic2016case}
Grgic-Hlaca, N.; Zafar, M.~B.; Gummadi, K.~P.; and Weller, A. 2016.
\newblock The case for process fairness in learning: Feature selection for fair
  decision making.
\newblock In \emph{NIPS symposium on machine learning and the law}, 2, 11.
  Barcelona, Spain.

\bibitem[{Hardt, Price, and Srebro(2016)}]{hardt_equality_2016}
Hardt, M.; Price, E.; and Srebro, N. 2016.
\newblock Equality of Opportunity in Supervised Learning.
\newblock In Lee, D.~D.; Sugiyama, M.; von Luxburg, U.; Guyon, I.; and Garnett,
  R., eds., \emph{Advances in Neural Information Processing Systems 29: Annual
  Conference on Neural Information Processing Systems 2016, December 5-10,
  2016, Barcelona, Spain}, 3315--3323.

\bibitem[{Klebanov(2014)}]{klebanov_precise_2014}
Klebanov, V. 2014.
\newblock Precise quantitative information flow analysis - a symbolic approach.
\newblock \emph{Theor. Comput. Sci.}, 538: 124--139.

\bibitem[{Kusner et~al.(2017)Kusner, Loftus, Russell, and
  Silva}]{kusner2017counterfactual}
Kusner, M.~J.; Loftus, J.~R.; Russell, C.; and Silva, R. 2017.
\newblock Counterfactual Fairness.
\newblock In Guyon, I.; von Luxburg, U.; Bengio, S.; Wallach, H.~M.; Fergus,
  R.; Vishwanathan, S. V.~N.; and Garnett, R., eds., \emph{Advances in Neural
  Information Processing Systems 30: Annual Conference on Neural Information
  Processing Systems 2017, December 4-9, 2017, Long Beach, CA, {USA}},
  4066--4076.

\bibitem[{Lampson(1973)}]{Lampson73}
Lampson, B.~W. 1973.
\newblock A Note on the Confinement Problem.
\newblock \emph{Commun. {ACM}}, 16(10): 613--615.

\bibitem[{Loftus et~al.(2018)Loftus, Russell, Kusner, and
  Silva}]{loftus_causal_algorithmic_fairness}
Loftus, J.~R.; Russell, C.; Kusner, M.~J.; and Silva, R. 2018.
\newblock Causal Reasoning for Algorithmic Fairness.
\newblock \emph{CoRR}, abs/1805.05859.

\bibitem[{Mitchell et~al.(2021)Mitchell, Potash, Barocas, D'Amour, and
  Lum}]{mitchell2018predictionbased}
Mitchell, S.; Potash, E.; Barocas, S.; D'Amour, A.; and Lum, K. 2021.
\newblock Algorithmic Fairness: Choices, Assumptions, and Definitions.
\newblock \emph{Annual Review of Statistics and Its Application}, 8(1):
  141--163.

\bibitem[{Pearl(2009)}]{pearl2009causal}
Pearl, J. 2009.
\newblock Causal inference in statistics: An overview.
\newblock \emph{Statistics Surveys}, 3: 96–146.

\bibitem[{Ramadan et~al.(2019)Ramadan, Ahmadian, J{\"{u}}rjens, Staab, and
  Str{\"{u}}ber}]{ramadan_explaining_2019}
Ramadan, Q.; Ahmadian, A.~S.; J{\"{u}}rjens, J.; Staab, S.; and Str{\"{u}}ber,
  D. 2019.
\newblock Explaining Algorithmic Decisions with respect to Fairness.
\newblock In Becker, S.; Bogicevic, I.; Herzwurm, G.; and Wagner, S., eds.,
  \emph{Software Engineering and Software Management, {SE/SWM} 2019, Stuttgart,
  Germany, February 18-22, 2019}, volume {P-292} of \emph{{LNI}}, 161--162.
  {GI}.

\bibitem[{Ramadan et~al.(2018)Ramadan, Ahmadian, Str{\"{u}}ber, J{\"{u}}rjens,
  and Staab}]{ramadan_model-based_2018}
Ramadan, Q.; Ahmadian, A.~S.; Str{\"{u}}ber, D.; J{\"{u}}rjens, J.; and Staab,
  S. 2018.
\newblock Model-based discrimination analysis: a position paper.
\newblock In Brun, Y.; Johnson, B.; and Meliou, A., eds., \emph{Proceedings of
  the International Workshop on Software Fairness, FairWare@ICSE 2018,
  Gothenburg, Sweden, May 29, 2018}, 22--28. New York, NY, USA: Association for
  Computing Machinery.

\bibitem[{Smith(2009)}]{Smith2009}
Smith, G. 2009.
\newblock On the Foundations of Quantitative Information Flow.
\newblock In de~Alfaro, L., ed., \emph{Foundations of Software Science and
  Computational Structures, 12th International Conference, {FOSSACS} 2009, Held
  as Part of the Joint European Conferences on Theory and Practice of Software,
  {ETAPS} 2009, York, UK, March 22-29, 2009. Proceedings}, volume 5504 of
  \emph{LNCS}, 288--302. Cham: Springer.

\bibitem[{Snelting et~al.(2014)Snelting, Giffhorn, Graf, Hammer, Hecker, Mohr,
  and Wasserrab}]{joana14it}
Snelting, G.; Giffhorn, D.; Graf, J.; Hammer, C.; Hecker, M.; Mohr, M.; and
  Wasserrab, D. 2014.
\newblock Checking Probabilistic Noninterference Using JOANA.
\newblock \emph{it - Information Technology}, 56: 280--287.

\bibitem[{Taskesen et~al.(2021)Taskesen, Blanchet, Kuhn, and
  Nguyen}]{taskesen_statistical_2021}
Taskesen, B.; Blanchet, J.~H.; Kuhn, D.; and Nguyen, V.~A. 2021.
\newblock A Statistical Test for Probabilistic Fairness.
\newblock In Elish, M.~C.; Isaac, W.; and Zemel, R.~S., eds., \emph{FAccT '21:
  2021 {ACM} Conference on Fairness, Accountability, and Transparency, Virtual
  Event / Toronto, Canada, March 3-10, 2021}, 648--665. New York, NY, USA:
  Association for Computing Machinery.

\bibitem[{Teuber and Beckert(2023{\natexlab{a}})}]{TeuberEWAF23}
Teuber, S.; and Beckert, B. 2023{\natexlab{a}}.
\newblock Formally Verified Algorithmic Fairness Using Information-Flow Tools.
\newblock In Alvarez, J.~M.; Fabris, A.; Heitz, C.; Hertweck, C.; Loi, M.; and
  Zehlike, M., eds., \emph{Proceedings of the 2nd European Workshop on
  Algorithmic Fairness, Winterthur, Switzerland, June 7th to 9th, 2023}, volume
  3442 of \emph{{CEUR} Workshop Proceedings}. CEUR-WS.org.

\bibitem[{Teuber and Beckert(2023{\natexlab{b}})}]{Artefact}
Teuber, S.; and Beckert, B. 2023{\natexlab{b}}.
\newblock samysweb/AAAI24-Fairness: AAAI24 Artefact.
\newblock {10.5281/zenodo.10385360}.

\bibitem[{Teuber and Weigl(2021)}]{teuber_quantifying_2021}
Teuber, S.; and Weigl, A. 2021.
\newblock Quantifying Software Reliability via Model-Counting.
\newblock In Abate, A.; and Marin, A., eds., \emph{Quantitative Evaluation of
  Systems - 18th International Conference, {QEST} 2021, Paris, France, August
  23-27, 2021, Proceedings}, volume 12846 of \emph{LNCS}, 59--79. Cham:
  Springer.

\bibitem[{Yang and Meel(2023)}]{DBLP:conf/cav/YangM23}
Yang, J.; and Meel, K.~S. 2023.
\newblock Rounding Meets Approximate Model Counting.
\newblock In Enea, C.; and Lal, A., eds., \emph{Computer Aided Verification -
  35th International Conference, {CAV} 2023, Paris, France, July 17-22, 2023,
  Proceedings, Part {II}}, volume 13965 of \emph{Lecture Notes in Computer
  Science}, 132--162. Springer.

\end{thebibliography}
